\documentclass[aps,pre,superscriptaddress,longbibliography,twocolumn]{revtex4-2}
\usepackage{graphicx}
\usepackage{epstopdf}
\usepackage{bm}
\usepackage{amsmath}
\usepackage{verbatim}
\usepackage{amssymb}
\usepackage{listings}
\usepackage{color}
\definecolor{darkblue}{rgb}{0,0,0.6}
\usepackage[colorlinks,linkcolor=darkblue,citecolor=darkblue,urlcolor=darkblue]{hyperref}
\usepackage{xspace}

\newcommand{\bigO}[1]{\ensuremath{\OCAL(#1)}}
\newcommand{\average}[1]{\left\langle #1 \right\rangle}

\newcommand{\UN}[1]{U^{#1}_{\infty}}
\newcommand{\UNCUT}[2]{U^{(#1)}_{#2}}

\newcommand{\SD}{\texttt{SoftDisks}\xspace}

\renewcommand\vec[1]{\boldsymbol{#1}}

\newcommand\subcap[1]{{(#1):}}

\newcommand{\eq}[1]{Eq.~\eqref{#1}}

\newcommand{\fig}[1]{Fig.~\ref{#1}}
\newcommand{\quot}[1]{``#1''}

\newcommand{\sect}[1]{Section~\ref{#1}}

\newcommand{\OCAL}{\mathcal{O}}  
\newcommand{\glb}{\left(}  
\newcommand{\grb}{\right)}  
\newcommand{\glc}{\left[}  
\newcommand{\grc}{\right]}  

\begin{document}

\title{The Liquid--Hexatic Transition for Soft Disks 
}

\author{Yoshihiko Nishikawa}
\affiliation{Graduate School of Information Sciences, Tohoku University, Sendai 
980-8579, Japan}

\author{Werner Krauth}
\affiliation{Laboratoire de Physique de l’Ecole normale sup\'erieure, ENS,
Universit\'e PSL, CNRS, Sorbonne Universit\'e, Universit\'e de Paris Cit\'e, 
Paris, France}

\author{A.~C.~Maggs}
\affiliation{CNRS UMR7083, ESPCI Paris, Universit\'e PSL, 10 rue
Vauquelin, 75005 Paris, France}

\date{\today} 

\begin{abstract}
We study the liquid--hexatic transition of soft disks with massively parallel
simulations and determine the equation of state as a function of system size.
For systems with interactions decaying as the inverse $m$th power of the
separation, the liquid--hexatic phase transition is continuous for $m = 12$ and
$m=8$, while it is of first order for $m = 24$. The critical power $m$ for the
transition between continuous and first-order behavior is larger than previously
reported. The continuous transition for  $ m=12 $ implies that the
two-dimensional Lennard-Jones model has a continuous liquid--hexatic transition
at high temperatures. We also study the Weeks--Chandler--Andersen model and find
a continuous transition at high temperatures, that is consistent with the
soft-disk case for $m=12$. Pressure data as well as our implementation are
available from an open-source repository.
\end{abstract}

\maketitle

\section{Introduction}

Two-dimensional melting transitions are observed  in multiple settings,
including adatoms on metal surfaces~\cite{Liu2005,Negulyaev2009}, colloids in
confined geometries~\cite{Kusner1994,Ebert2009,ThorneyworkDullens2017},
skyrmions in magnetic materials~\cite{Nishikawa2019,Reichhardt2022}, and trapped
electrons on the surface of $^4$He~\cite{Grimes1979,Glattli1988}.  Unlike their
three-dimensional counterparts, that generically feature first-order
liquid--crystal transitions, the nature of the phase transitions of
two-dimensional particle systems depends on the details of interaction
potentials and particle shapes~\cite{Kapfer2015PRL,Anderson2017,Li2020}.  In
systems with short-range interactions, crystalline phases with long-range
density correlations do not exist, since phonon excitations imply diverging
fluctuations~\cite{Peierls1935,Mermin1968,Richthammer}. This mirrors the physics
of two-dimensional $O(n)$ spin models, where phase transitions are absent for $n
\geq 3$~\cite{Mermin1966}. For $n=2$ (the XY model) the low-temperature phase
behavior is characterized by power-law spin--spin correlations and the presence
of pairs of topological vortices. The Kosterlitz--Thouless phase transition
between the two phases is now solidly
established~\cite{KosterlitzThouless1973,FroehlichSpencer1981,HasenbuschXY2005,
Kosterlitz_2016}.

Particle systems in two dimensions  sustain two types of topological
defect~\cite{KosterlitzThouless1973,Nelson1978,Nelson1979,
Young1979VectorCoulomb}, disclinations and dislocations. The KTHNY theory
proposes that high-density solids melt into a liquid via two successive
Kosterlitz--Thouless transitions, corresponding to the successive unbinding of
dislocations, and then of disclinations. Between these two transitions, the
intermediate hexatic phase 
short-range positional order and quasi-long-range orientational order. In an
alternative scenario~\cite{Fisher1979,Chui1983}, the solid melts directly into a
liquid via a first-order transition due to the formation of grain boundaries.
Most of the theories of two-dimensional melting worked within these two
frameworks. For the special case of hard disks, however, decades of numerical
studies going back to the dawn of Monte Carlo and molecular dynamics
simulations~\cite{Metropolis1953,Alder1962,Li2022hard}, finally
concluded~\cite{Bernard2011} that nature chooses  a first-order liquid--hexatic
transition, and a continuous Kosterlitz--Thouless hexatic--solid transition.
These results were confirmed in recent
experiments~\cite{ThorneyworkDullens2017}, and they contradict the historic
scenarios~\cite{Bernard2011,Engel2013}.

The nature of the melting transition of soft disks with, for example,
Lennard-Jones or (inverse) power-law potentials, may depend on
model parameters, such as temperature and density. In Lennard-Jones systems, a
number of conflicting transition scenarios have been
reported~\cite{Strandburg1984,Chen1995,Wierschem2011,
Hajibabaei2019,Toledano2021,Li2020,Li2020b}. The Lennard-Jones phase 
diagram can also be related to the case of power-law potentials of the form 
$U(r) = {(\sigma / r)}^{m}$~\cite{Kapfer2015PRL,Hajibabaei2019,Toledano2021}, 
which interpolate between hard disks, $m=\infty$, and the soft potential with $m=3$,
for which the KTHNY scenario is well-established~\cite{Zahn1999}.

The debate and controversies as to the order of the liquid--hexatic transition
in two-dimensional soft-disk systems are due to remarkably strong finite-size
effects. The determination of the order of the transition has largely relied on
the presence or absence of a Mayer--Wood loop in the equation of
state~\cite{MayerWood1965,Kapfer2015PRL,Toledano2021,Anderson2017}. The major
difficulty is that the equation of state of finite systems may have a
Mayer--Wood loop even when a transition is continuous. The loop then vanishes at
a very large system size~\cite{Alonso1999}. It is thus impossible to conclude
on the nature of the transition with simulations on a single system size.
Careful analysis of finite-size effects must be performed to reach a definitive
conclusion.

In this paper, we revisit the liquid--hexatic transition for soft disks with
large-scale parallel algorithms implemented on high-performance Graphics
Processing Units (GPU). We generate high-precision Monte Carlo data on multiple
system sizes up to $N=1024^2$ particles to better distinguish between the
different scenarios. Following Ref.~\cite{AndersonGPU2013},  we implement a
massively parallel Metropolis algorithm, and determine equations of state to
high precision for the power-law potentials with $m=8$, $12$, $24$, together
with the  Weeks--Chandler--Andersen (WCA) potential~\cite{Weeks1971} at two
different temperatures. We  focus on the equation of state and carefully analyze
the finite-size scaling of the free-energy barrier to determine  the nature of
the liquid--hexatic transition.

This paper is organized as follows. In \sect{sec:models}, we introduce the
potentials that we study, together with the observables that we measure. We
present our Monte Carlo results for inverse power-law models in \sect{sec:IPL}
and for the WCA model in \sect{sec:WCA}. Using the equation of state, we perform
a detailed finite-size analysis and discuss the liquid--hexatic transition. In
\sect{sec:summary}, we summarize the results and present our conclusions as to
the order of the liquid--hexatic transition.

\section{Models and method}
\label{sec:models}

We consider $N$ disks in a square periodic box of volume $V$,
with two
different interaction potentials. We are interested in the generic soft-disk
model
\begin{equation}
\UN{m}(r) / \epsilon =   {(r / \sigma)} ^{-m} \quad (r > 0), 
\label{eq:SoftNoCutoff}
\end{equation}
where $\sigma$, a diameter, provides a length scale. Its phase behavior
depends on the single parameter $\Gamma = \beta \epsilon {(\phi\sigma)}^{n/2}$,
where $\beta$ is the inverse temperature and $\phi = N / V$ is the number
density.  Soft-disk systems with the infinite-range potential $\UN{m}$ can be
simulated with a computational complexity of \bigO{1} per move (that is per
event) using the cell-veto event-chain Monte Carlo
algorithm~\cite{KapferKrauth2016}. Nevertheless, for computational convenience,
a cutoff $r_c$ is often introduced in studies of long-range
potentials~\cite{Smit1991,Smit1992}.  Simply restricting $ \UN{m}$ to
separations smaller than a value $r_c$ complicates the calculation of virial
pressures because of the discontinuity~\cite{Wittmer2013,Wittmer2013b} in the
potential at $r=r_c$.  These corrections in the pressure are absent in a
continuous, shifted potential
\begin{equation}
\UNCUT{m}{r_c}(r) / \epsilon = 
\begin{cases}
\glb r / \sigma  \grb ^{-m} - \glb  r_c /  \sigma \grb ^{-m} & 
(r < r_c)\\
0 & (r > r_c).
\end{cases}
\label{eq:SoftShifted}
\end{equation}
This potential was studied in Ref.~\cite{KapferKrauth2016} with a cutoff
$r_c/\sigma = 1.8$, which we again use in this paper. The exact scaling in
temperature and density, via $\Gamma$, does not strictly hold with a cutoff.
However, we change only the number density and fix $\beta\epsilon = 1$  for this
family of potentials (see \sect{sec:IPL} for a discussion of the effects of the
cutoff on the melting).

Our second model is the
Weeks--Chandler--Anderson (WCA) model, that features a
Lennard-Jones potential with a cutoff  at its
minimum shifted for continuity~\cite{Weeks1971}:
\begin{equation}
U_\text{WCA}(r) / \epsilon =
\begin{cases}
4 \glc \glb\frac{\sigma}{r} \grb^{12} - \glb \frac{\sigma}{r} \grb ^6 \grc + 1, 
& \frac{r}{\sigma} < 2^{1/6}\\
0 & \frac{r}{\sigma} > 2^{1/6}.
\end{cases}
\label{eq:WCA}
\end{equation}
The WCA potential is purely repulsive (the additional factor of $4$ in
\eq{eq:WCA} compared to \eq{eq:SoftShifted} comes from the usual definition of
the Lennard-Jones model and is not taken into account in our
discussions). We study it as a function of
temperature and of number density. The system was studied previously at
$\beta\epsilon=1$, where it has a first-order liquid--hexatic
transition~\cite{Khali2021}. Because of their identical potential for small
$r/\sigma$, the WCA model, the Lennard-Jones model, and the $m=12$
model have the same phase behavior at high temperatures and pressures.

We have implemented a massively parallel Metropolis algorithm on
GPUs~\cite{AndersonGPU2013,Li2022hard}.
Our implementation reaches
$1.7\times 10^{13}$ individual Monte Carlo trials per hour for the power-law
model with $m=12$ with $N=1024^2$ on an NVIDIA GeForce RTX 3090 GPU, which is
more than $1000$ times faster than a sequential implementation. Our code
thus requires some $4$ days for $1.6\times 10^9$ sweeps for $1024^2$ disks.

\begin{figure}[t]
\includegraphics[width=\linewidth]{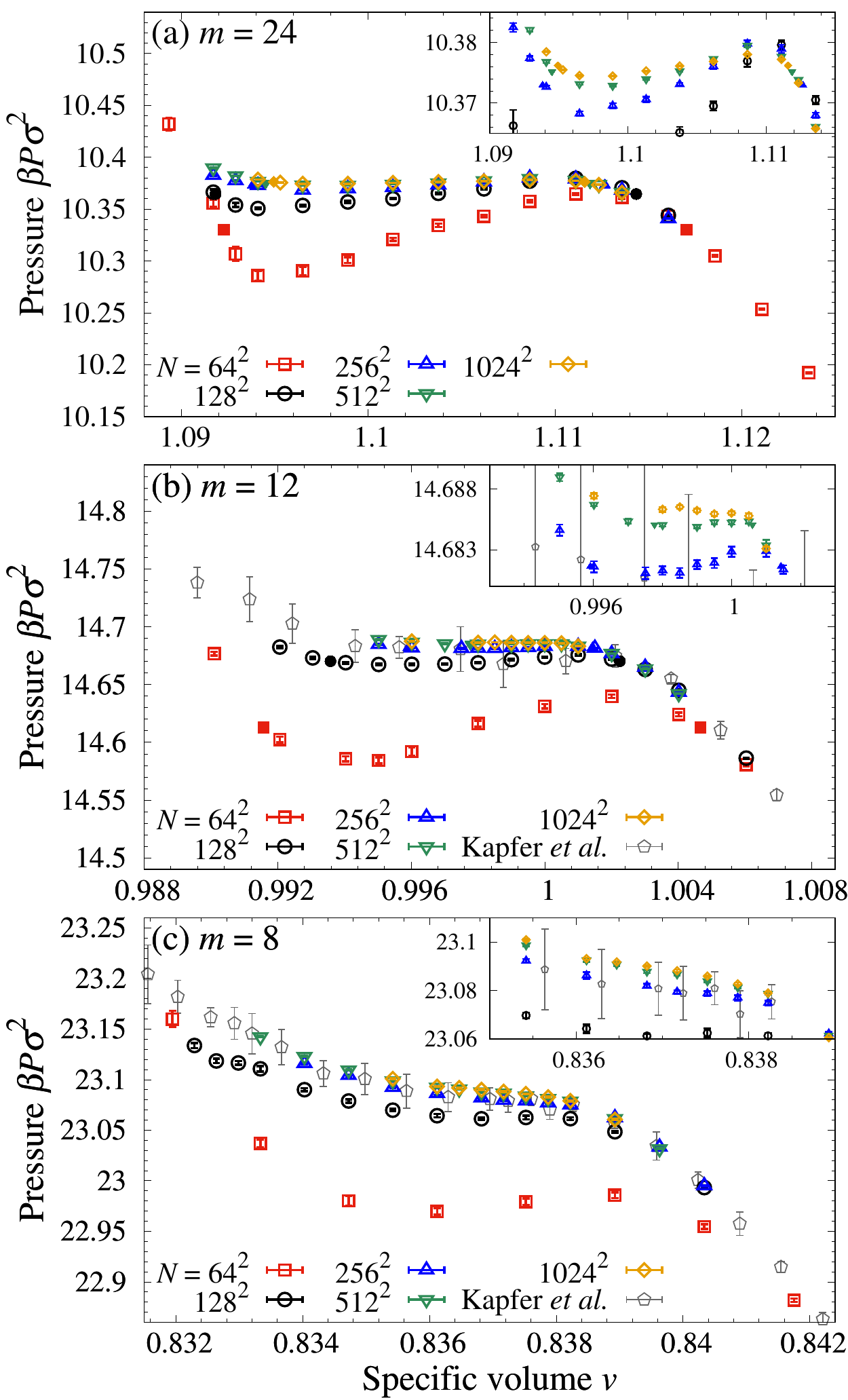}
\caption{Equation of state of the power-law model of \eq{eq:SoftShifted} for
$N=64^2$, $128^2$,
$256^2$, $512^2$, and  $1024^2$. \subcap{a} For $m=24$, the loop is pronounced
for all system sizes. \subcap{b} For $m=12$, the loop is tiny for $N=512^2$ and
$1024^2$ and is expected to disappear for larger $N$. \subcap{c} $m=8$. The loop
disappears for $N=256^2$, and the transition is clearly continuous. Data labeled
as  \quot{Kapfer \textit{et al.}}~\cite{Kapfer2015PRL} is for $N\approx 255^2$.
Filled symbols for each system size indicate the
specific volumes of the liquid and hexatic phases, $v_\text{liq}$ and $v_\text{hex}$ 
($v_\text{liq} > v_\text{hex}$).
}
\label{fig:eqstate_ipl}
\end{figure}

We measure the equilibrium pressure  from the virial
\begin{equation}
P = \frac{\phi}{\beta} - \frac1{2V} 
\average{\sum_{i>j} r_{ij}\frac{\partial U(r_{ij})}{\partial r_{ij}}},
\label{eq:pressure}
\end{equation}
where $r_{ij} = |\vec r_i - \vec r_j|$. For each system, we produce a single
time series for the pressure, and estimate its correlations
using the stationary bootstrap
method~\cite{Politis1994,Politis2004,Patton2009,nishikawa2021stationary}.

For hard disks as well as soft disks with large $m$, the equation of state has
a `loop', that is a non-monotonic variation of the pressure as a function of
(inverse) density, close to the liquid--hexatic phase transition. The equation
of state then becomes flat over a finite range of inverse density when $N\to
\infty$.  The existence of a loop in the equation of state of a finite system
has been taken to indicate a first-order transition. However, the loop does not
necessarily mean a first-order transition~\cite{MayerWood1965,Binder2012} as we
will show below. In order to determine the nature of the transition, it is
essential to observe the finite-size dependence of the equation of state.

The Mayer--Wood loop in the equation of state (with the pressure being the
derivative of the free energy with respect to the volume) results from a
non-convex free energy as a function of the volume.
In the presence of a first-order transition, that is, of coexistence of two
phases of two distinct specific volumes,
the free energy $F(v)$ as
a function of specific volume $v = {(\phi \sigma^2)}^{-1}$ has two minima
separated by a free-energy barrier $\Delta F$. Then the free-energy barrier
scales as $\Delta F = O(L^{d-1})$ goes to $0$ and $f(v) = F(v) / N$ is convex
when $N\to \infty$.

Nevertheless, $\Delta f$ can be nonzero and positive in
finite systems~\cite{Alonso1999}. The scaling of $\Delta f$ as a function of
$N$ depends on the nature of the transition. For a first-order transition, the
barrier comes from the surface free energy of a single compact droplet and thus
$\Delta f = O(L^{d-1} / N) = O(L^{-1})$ in spatial dimension $d$, where $L$ is
the size of the simulation box. The free-energy barrier $\Delta f$ should decay
faster than $1/L$ for a continuous transition. We expect that the free energy 
becomes strictly convex and $\Delta f=0$ at large finite $N$ for 
a continuous liquid--hexatic transition.
Practically, we numerically integrate the equation of state (as a function of
the specific volume) to obtain the free energy and the specific 
volumes for the liquid and hexatic phases, $v_\text{liq}$ and 
$v_\text{hex}$ respectively, using a spline interpolation,
and the bootstrap method to estimate the errors of $\Delta f$,
$v_\text{liq}$, and $v_\text{hex}$.

\section{Power-law models}
\label{sec:IPL}

\begin{figure}[t]
\includegraphics[width=\linewidth]{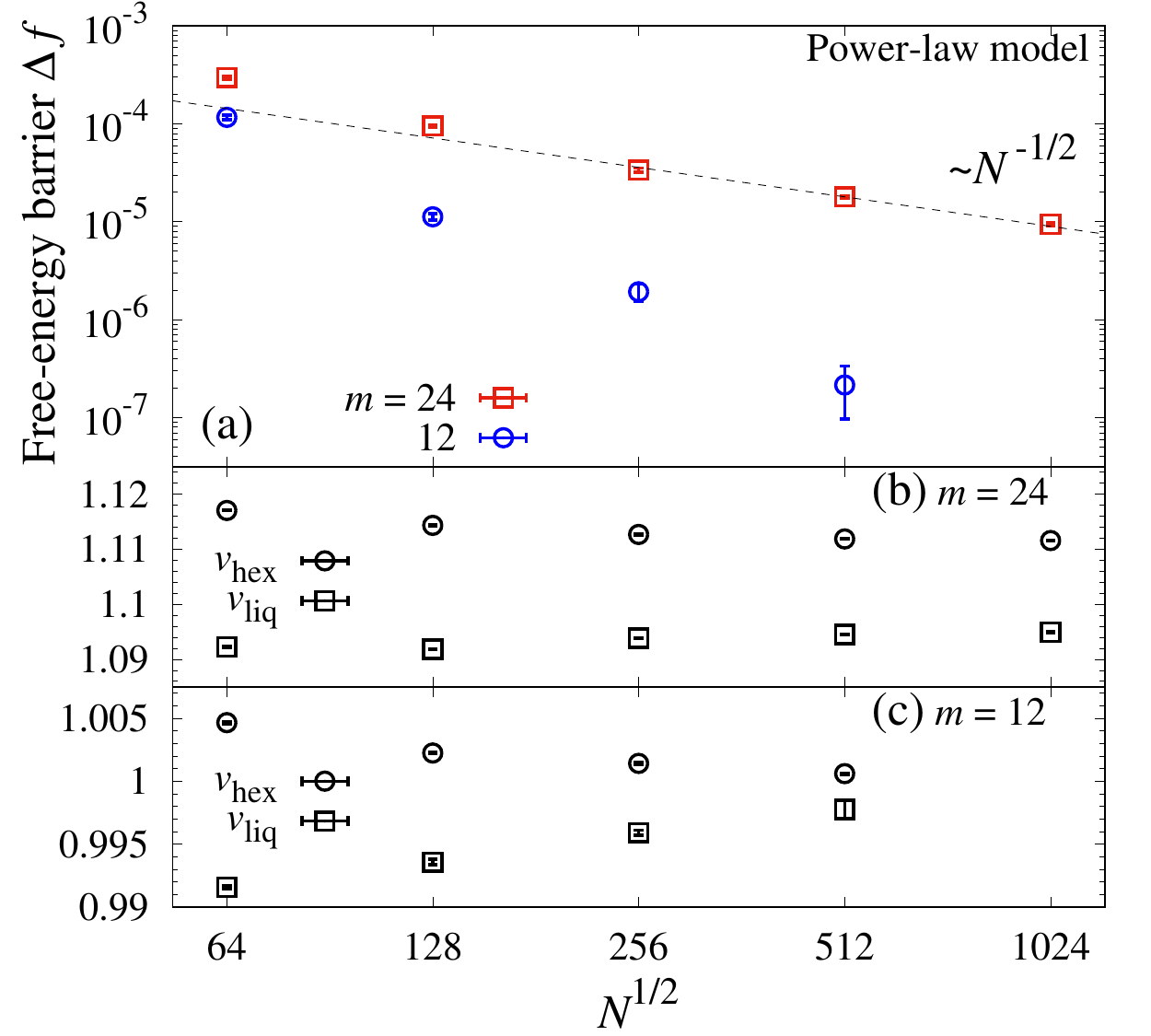}
\caption{\subcap{a} Free-energy barrier $\Delta f$ as a function of $N^{1/2}$ for the
power-law model of \eq{eq:SoftShifted} with $m=24$ and $m=12$. For $m=24$,
$\Delta f$ asymptotically scales as $N^{-1/2}$, indicating a first-order
transition. For $m=12$, it decays faster, indicating a continuous transition. 
Specific volumes $v_\text{liq}$ and $v_\text{hex}$ as functions of $N^{1/2}$, 
for \subcap{b} $m=24$ and \subcap{c} $m=12$.
For $m=24$, the specific volumes are well separated, consistent with the first-order 
transition. For $m=12$, on the other hand, $v_\text{liq}$ and $v_\text{hex}$ approach 
each other and eventually merge at $N=1024^2$, as seen in the monotonic equation 
of state \fig{fig:eqstate_ipl}~(b).
}
\label{fig:barrier_ipl}
\end{figure}

In this section, we study soft disks with potential
$\UNCUT{m}{r_c}$ for $m=24$, $12$, and $8$. The equilibrium pressure of the
system is computed using \eq{eq:pressure}. We show in \fig{fig:eqstate_ipl}
the equation of state, plotting the dimensionless
pressure $\beta P\sigma^2$ as a function of specific volume $v =
{(\phi\sigma^2)}^{-1}$.

For $m=24$, we observe a Mayer--Wood loop for small
numbers of disks. While the loop amplitude decreases with $N$, it
survives up to $N=1024^2$,  \fig{fig:eqstate_ipl}~(a). A similar $N$
dependence of the equation of state has been observed in hard disks, equivalent to the $m\to\infty$ limit of the inverse power 
law~\cite{Bernard2011,Engel2013,Li2022hard}.  We quantify the decay of 
the loop
amplitude by the free-energy barrier $\Delta f$,  \fig{fig:barrier_ipl}~(a).
$\Delta f$ decreases with increasing $N$.
The barrier $\Delta f$ 
asymptotically scales as $\Delta f \sim N^{-1/2}\sim 1/L$ for large $N$, similarly to
hard disks~\cite{Bernard2011}. This is a strong indication of a
first-order transition. The specific volumes of the liquid and hexatic phases
$v_\text{liq}$ and $v_\text{hex}$ are well separated and do not merge,
meaning the coexisting phase over a finite range of specific volume, 
see \fig{fig:barrier_ipl}~(b).

The case $m=12$ also has a clear
loop in the equation of state from $v \simeq 1.005$ to $0.992$ when
$N=64^2$,  \fig{fig:eqstate_ipl}~(b). 
In Ref.~\cite{Kapfer2015PRL} it was concluded that the liquid--hexatic
transition is first order.
However, with increasing $N$, the loop shrinks too rapidly. 
Note that for $N=256^2$,
our Monte Carlo results are consistent with Ref.~\cite{Kapfer2015PRL}
for  $N=6.5\times 10^4 \approx 255^2$.
$\Delta f$ decays faster than the scaling $\Delta
f \sim N^{-1/2}$ expected when the transition is of
first order, and $v_\text{liq}$ and $v_\text{hex}$
merge at $N \simeq 1024^2$, see \fig{fig:barrier_ipl}~(a) and (c).
We thus conclude
that for $m=12$ the
liquid--hexatic transition is continuous, so that
the hexatic melts into the liquid via a
Kosterlitz--Thouless transition, following the conventional two-step
scenario
~\cite{HalperinNelson1978,Nelson1978,Nelson1979,Young1979VectorCoulomb}.

We observe similar finite-size effects for $m=8$ where the
equation of state of the system for $N=128^2$ is almost monotonic; the
continuous nature of the liquid--hexatic transition is clear.
The conclusion drawn in Ref.~\cite{Kapfer2015PRL} was thus correct
qualitatively but not quantitatively: The nature of the liquid--hexatic
transition indeed changes from first-order to continuous at finite $m$, but not
at $m\lesssim 6$. Our results show that the critical value of $m_c$ separating
the two regimes lies between $m=12$ and $24$, and the $m=12$ model is already
in the regime of a continuous liquid--hexatic transition as pointed out in
Ref.~\cite{Hajibabaei2019}. Nevertheless, the system size
$N\gtrsim 512^2$ to have the equation of state monotonic for $m=12$ is significantly
larger than $N\simeq 128^2$ for $m=8$ model. 
We expect this system size to have the equation of state monotonic depends on $m$
and grows approaching the critical $m_c$ from below, eventually diverging at $m_c$.

We now comment on the effect of the cutoff on the 
transition.  When the cutoff  $r_c$ is too small, the model defined by the
interaction potential \eq{eq:SoftShifted} behaves differently from the original
power-law model and the melting scenario could change. When $m$
is large the interaction potential is shifted by only a small
amount. With $r_c = 1.8\sigma$, the shift in \eq{eq:SoftShifted} is
${(\sigma / r_c)}^{24} < 10^{-6}$.  However, the shifts for $m=12$ and
$m=8$  are larger. We  estimate the change in pressure due to the
cutoff in a single phase system as
\begin{equation}
  \Delta P = + \frac{\phi^2}{4} \int_{r_c}^\infty
  r \frac{d\UN{m}(r)}{dr}\,
2\pi  g(r)\,
 r dr,
\end{equation}
where $g(r)$ is the radial distribution function of the model with interaction
potential $\UN{m}(r)$.  If $g(r) \sim 1$ beyond the cutoff, then
$\Delta P \sim -\phi^2$. In the case of a monotonically decreasing equation of
state, this correction lowers the pressure most strongly for small
$\phi^{-1}$, on the left of Fig.~\ref{fig:eqstate_ipl}, pushing a transition towards first order.  We confirmed, for $m=12$,
that a cutoff at $r_c = 1.12\sigma$ enhances the loop in the equation of state.
For $r_c = 1.8\sigma$, the liquid--hexatic transition is continuous for $m=12$
and $8$, should not change our predictions as to the nature of the transition.

\section{WCA model}
\label{sec:WCA}

\begin{figure}[t]
\includegraphics[width=\linewidth]{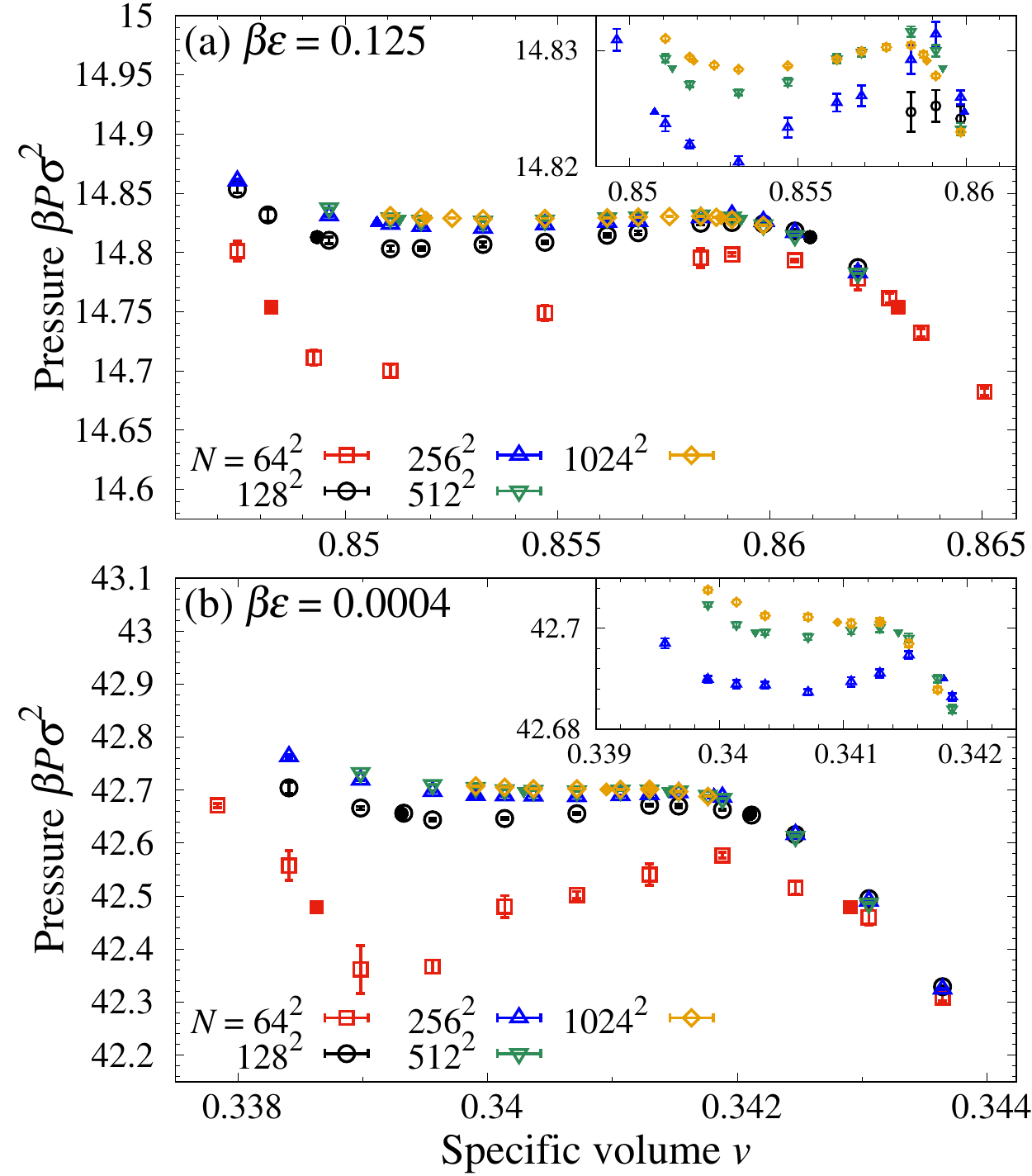}
\caption{Equation of state of the WCA model at (a) $\beta\epsilon=0.125$ and (b)
$\beta\epsilon=0.0004$.  Each inset shows the equation of state close to  the
liquid--hexatic transition. Filled symbols for each system size indicate the
specific volumes of the liquid and hexatic phases, $v_\text{liq}$ and $v_\text{hex}$
($v_\text{liq} > v_\text{hex}$).
}
\label{fig:eqstate_wca}
\end{figure}

\begin{figure}[t]
\includegraphics[width=\linewidth]{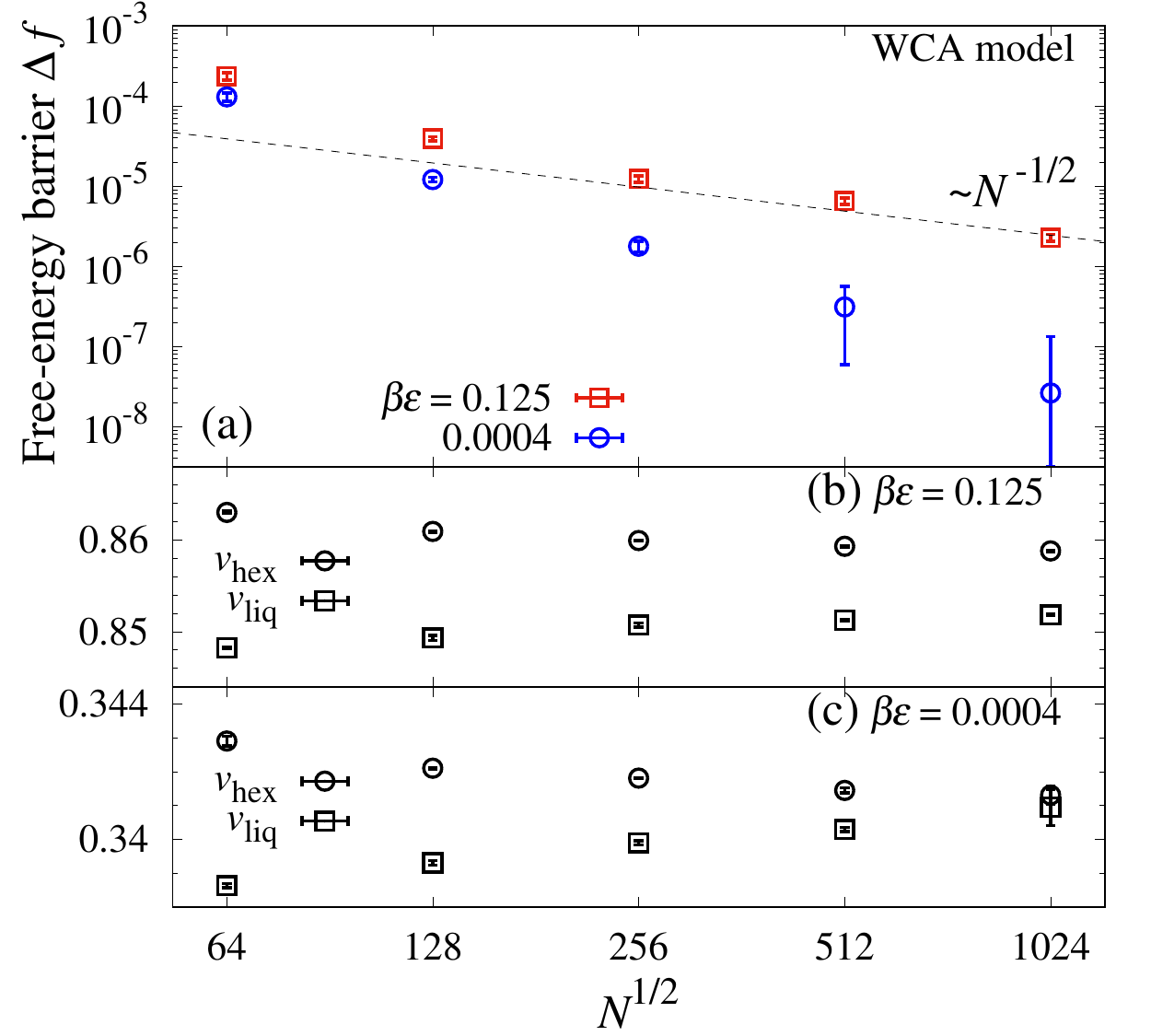}
\caption{Free-energy barrier $\Delta f$ as a function of $N^{1/2}$ for
the WCA model at $\beta \epsilon=0.125$ and $\beta \epsilon=0.0004$.  At
$\beta\epsilon=0.125$, the free-energy barrier $\Delta f$ crosses over to 
scaling in $N^{-1/2}$ for large $N$, suggesting a first-order transition.
$\Delta f$ decays  faster at $\beta \epsilon = 0.0004$, implying that the transition is
continuous.
Specific volumes $v_\text{liq}$ and $v_\text{hex}$ as functions of $N^{1/2}$, 
at \subcap{b} $\beta\epsilon = 0.125$ and \subcap{c} $\beta\epsilon = 0.0004$.
At $\beta\epsilon = 0.125$, the specific volumes do not merge
while at $\beta\epsilon = 0.0004$, they converge to almost the same value
at $N=1024^2$.
}
\label{fig:barrier_wca}
\end{figure}

In \sect{sec:IPL}, we confirmed that the $m=12$ power-law potential has a
continuous liquid--hexatic transition. Because the Lennard-Jones model has the
same behavior at small separations, it must obey the same continuous
liquid--hexatic scenario  at high enough temperatures. However, it was recently
claimed that the transition is first-order at all
temperatures~\cite{Li2020,Li2020b}.  In this section, we investigate this point
with the help of the truncated Lennard-Jones model~\eq{eq:WCA} which, at high
temperature or high density, is equivalent to the Lennard-Jones and the $m=12$
power-law model. We study the WCA model at $\beta\epsilon= 0.125$ and $0.0004$.
At low temperatures, effects due to the truncation should be strong. We expect
that the small value of the cutoff pushes the system to a first-order
liquid--hexatic transition. At high temperatures, the effect of truncation is
negligible and the WCA model should agree with what is observed in the power-law
model for $m=12$.

At low temperature $\beta\epsilon = 0.125$, the WCA equation of state
features a clear Mayer--Wood loop (see \fig{fig:eqstate_wca}~(a)), and the
amplitude decreases with increasing system size. The free-energy
barrier $\Delta f$ decays faster than the scaling of the first-order
transition, $\Delta f \sim N^{-1/2}$, when $N$ is small, but it approaches
this scaling when $N \gtrsim 256^2$, \fig{fig:barrier_wca}~(a) 
(see also~\cite{Khali2021}, at $\beta\epsilon=1$). The specific volumes 
$v_\text{liq}$ and $v_\text{hex}$ are well separated (see \fig{fig:barrier_wca}~(b)), 
and the coexisting phase appears between $v_\text{liq} \simeq 0.86$ and 
$v_\text{hex} \simeq 0.85$, confirming a first-order transition of 
the WCA model at low temperature.

At high temperature $\beta\epsilon = 0.0004$, the WCA-model equation of state
features a loop at $0.34 \lesssim v \lesssim 0.342$ for small system
sizes, but its amplitude decreases rapidly, and it vanishes
for $N\gtrsim 512^2$,  \fig{fig:eqstate_wca}~(b).  
The free-energy barrier $\Delta f$ decays
faster than the first-order scaling $\Delta f \sim N^{-1/2}$.
Consistently, $v_\text{liq}$ and $v_\text{hex}$ merges at $N\simeq 1024^2$,
meaning that the system does not have a coexisting phase, see \fig{fig:barrier_wca}~(c).
We thus conclude that, at high enough temperatures, the transition in 
the WCA model becomes
continuous as expected from our results of the $m=12$ model, contrary to the 
phase diagram shown in Refs.~\cite{Li2020,Li2020b}.

For the WCA model, the critical inverse temperature $\beta_c$
separating the continuous and first-order liquid--hexatic transitions lies
in the range $\beta_c \epsilon  \in (0.0004, 0.125)$.
These results are consistent with Refs.~\cite{Li2020,Li2020b} who, for the
Lennard-Jones model, did not observe a continuous transition
at inverse temperature $\beta \epsilon = 10$,
but are inconsistent with $\beta_c \epsilon \simeq 0.91$
claimed in Refs.~\cite{Hajibabaei2019,Toledano2021},

\section{Summary and Discussion}
\label{sec:summary}

We have studied the equation of state of two-dimensional soft disks with
power-law interactions and for the WCA model, with its specifically truncated
Lennard-Jones interaction. A massively parallel Metropolis algorithm allowed us
to study up to $N=1024^2$ disks at  densities around the liquid--hexatic phase
transition and to estimate the equilibrium pressure with high precision.  We
identified the nature of the phase transition by the scaling of the free-energy
barrier with system size. Our results show that the order of the phase
transition depends on the exact form of the potentials and, for the WCA model,
on temperature.
We found that the power-law model with 
$m=24$ has a first-order
phase transition, consistent with the previous study~\cite{Kapfer2015PRL}.
For softer interactions, $m=12$ and $m=8$, however, the phase transition is
continuous.  Whereas we find easily that the $m=8$ model has a continuous
transition from the monotonic equation of state, the $m=12$ model requires
a careful analysis of the system-size dependence to demonstrate its continuous
nature: This model has a tiny but clear loop up to $N\simeq 256^2$, and
the equation of state becomes strictly monotonic
only for  $N\gtrsim 512^2$. The continuous nature of the $m=12$ model
is consistent with the conclusion in Ref.~\cite{Hajibabaei2019}.
The difference between the cases $m=24$ and $m=12$ clearly appears
in the system-size dependence of the free-energy barrier $\Delta f$.
We conclude that the parameter $m$ of
the power-law model changes
the nature of the transition, but the critical value of $m$ separating the two
regimes is between $m=12$ and $24$, not $m\lesssim 6$ reported
previously~\cite{Kapfer2015PRL}.

In the WCA model, temperature plays an analogous role to $m$ in the 
power-law model in changing the nature of the phase transition. At low
temperatures, as for  $\beta\epsilon = 0.125$,
the scaling of the free-energy barrier $\Delta f$ indicates a first-order transition.
At  high temperature, on the other hand, the transition becomes
continuous, with
finite-size effects that resemble those of the power-law model with
$m=12$.
A remaining puzzle is the value of the critical temperature $T_c$ separating
the
first-order and continuous liquid--hexatic transitions: Our Monte Carlo data on
the WCA model indicate
the critical transition between $\beta\epsilon= 0.0004$ and $0.125$, while
in the
conventional Lennard-Jones model it was claimed to be around $\beta_c\epsilon \simeq 0.91$~\cite{Hajibabaei2019,Toledano2021},
in disagreement with Ref.~\cite{Li2020b} which states that conventional Lennard-Jones at $\beta\epsilon=0.1$ has a first-order liquid--hexatic
transition. This could be either because the truncation in the WCA model
shifts the critical temperature, or because tiny but finite loop amplitudes 
at high temperatures were missed due to statistical noise in
Refs.~\cite{Hajibabaei2019,Toledano2021}. We will address this issue in future work.

The controversy over the two-dimensional melting transition has
originated in two difficulties.
First, the
very long timescales needed to reach equilibrium at high density and low temperature,
and secondly, strong finite-size effects in the equation of state. The
former difficulty
was partially resolved through event-chain Monte Carlo and massively parallel 
Metropolis algorithms. The latter, however, is still a source of difficulty.
The strong finite-size effects usually originate from a large length scale.
We expect the orientational correlation length should not be responsible 
although it diverges at the continuous liquid--hexatic transition. We have 
not identified this length scale yet, and will study this length scale in future 
work.

\acknowledgments{We thank S.~C.~Kapfer for providing the data presented in
Ref.~\cite{Kapfer2015PRL}. This research was conducted within the context of 
the International Research Project ``\textit{Non-Reversible Markov chains, Implementations 
and Applications}''. Y.N. is grateful for the kind support from Institut
Philippe Meyer and acknowledges support from JSPS KAKENHI (Grant No.~22K13968).
W.K. acknowledges support from the Alexander von Humboldt Foundation.}

\appendix

\section{Software package: outline, license, access}
\label{app:soft}

\normalsize

The present paper is accompanied by the \SD{} data and software package, which is 
published as an open-source project under the GNU GPLv3 license. \SD{} is 
available on GitHub as part of the  \texttt{JeLLyFysh} organization \footnote{The url of repository is \url{https://github.com/jellyfysh/SoftDisks}.}.  
The package contains a C++ program, using Nvidia GPU extensions, which implements 
the massively parallel algorithm of soft-disk model. This 
program is accompanied by Python scripts for the analysis of data.
The package also provides original 
pressure data for equations of state from Ref.~\cite{Kapfer2015PRL}, 
together with those obtained in the present work.

The \SD{} software package 
follows Ref.~\cite{AndersonGPU2013}. We overlay the full 2D system with a
four-color checkerboard of cells, each containing a small number of
disks. Different cells of the same color are separated by more than the
cutoff of the potential so that the Metropolis algorithm runs
independently on each of them. Monte Carlo trials that move a disk out of its cell
are rejected. After a fixed number of cycles (typically $16$), the cell system 
is displaced randomly. This permits disks to eventually move throughout the
system, as required for irreducibility. During simulations, we
measure physical quantities, such as the energy and
the pressure, which are calculated as sums within the local environment. We
also store snapshots of the system for later analysis.
Our GPU-based code will be useful for further study of
two-dimensional melting.

The Python scripts read the data from a simulation, and
perform a detailed analysis of thermodynamic properties.
Snapshots are used to generate detailed
movies of the time evolution of the system. Our analysis code
calculates spatial correlations and Voronoi tessellations using
NumPy and SciPy.
Equilibrium configurations obtained in this work are available
from \url{https://doi.org/10.5281/zenodo.7844567}.

\bibliography{refs.bib}

\end{document}